\documentclass[useAMS,usenatbib,fleqn]{mn2e}
\usepackage{graphicx}
\usepackage{epsfig}
\usepackage{color}
\usepackage{tabularx}
\usepackage{multirow}
\usepackage[para,online,flushleft]{threeparttable}
\usepackage[ampersand]{easylist}
\usepackage{amsmath}
\usepackage{hyperref}
\hypersetup{
    colorlinks=true,
    linkcolor=blue,
    filecolor=blue,      
    urlcolor=blue,
    citecolor=blue,
    filecolor=blue,
}
\title[Discovery of RP from SNR HB9]{Discovery of recombining plasma inside the extended gamma-ray supernova remnant HB9}
\author[Sezer et al.]{A.~Sezer,$^{1}$\thanks{E-mail: {\color {blue}aytap.sezer@avrasya.edu.tr} (AS)}
T.~Ergin,$^{2}$\thanks{{\color {blue}ergin.tulun@gmail.com} (TE)} R.~Yamazaki,$^{3}$\thanks{{\color {blue}ryo@phys.aoyama.ac.jp} (RY)} H.~Sano$^{4,5}$\thanks{{\color {blue}sano@a.phys.nagoya-u.ac.jp} (HS)} and Y.~Fukui$^{4,5}$\thanks{{\color {blue}fukui@a.phys.nagoya-u.ac.jp} (YF)} \\
$^{1}$Department of Electrical-Electronics Engineering, Avrasya University, 61250, Trabzon, Turkey\\
$^{2}$TUBITAK Space Technologies Research Institute, ODTU Campus, 06800, Ankara, Turkey\\
$^{3}$Department of Physics and Mathematics, Aoyama Gakuin University, 5-10-1 Fuchinobe, Sagamihara 252-5258, Japan\\
$^{4}$Institute for Advanced Research, Nagoya University, Furo-cho, Chikusa-ku, Nagoya 464-8601, Japan\\ $^{5}$Department of Physics, Nagoya University, Furo-cho, Chikusa-ku, Nagoya 464-8601, Japan
}
\begin{document}
\date{}
\pagerange{\pageref{firstpage}--\pageref{lastpage}} \pubyear{2019}
\maketitle
\label{firstpage}

\begin{abstract}
We present the results from the {\it Suzaku} X-ray Imaging Spectrometer observation of the mixed-morphology supernova remnant (SNR) HB9 (G160.9+2.6). We discovered recombining plasma (RP) in the  western {\it Suzaku} observation region and the spectra here are well described by a model having collisional ionization equilibrium (CIE) and RP components. On the other hand, the X-ray spectra from the eastern {\it Suzaku} observation region are best reproduced by the CIE and non-equilibrium ionization model. We discuss possible scenarios to explain the origin of the  RP emission based on the observational properties and concluded that the rarefaction scenario is a possible explanation for the existence of RP. In addition, the gamma-ray emission morphology and spectrum within the energy range of 0.2$-$300 GeV are investigated using $\sim$10 years of data from the {\it Fermi} Large Area Telescope (LAT). The gamma-ray morphology of HB9 is best described by the spatial template of radio continuum emission. The spectrum is well-fit to a log-parabola function and its detection significance was found to be $\sim$25$\sigma$. Moreover, a new gamma-ray point source located just outside the south-east region of the SNR's shell was detected with a significance of $\sim$6$\sigma$. We also investigated the archival H\,{\sc i} and CO data and detected an expanding shell structure in the velocity range of $-10.5$ and $+1.8$ km s$^{-1}$ that is coinciding with a region of gamma-ray enhancement at the southern rim of the HB9 shell.
\end{abstract}


\section{Introduction}
Thermal composite or mixed-morphology supernova remnants (MM SNRs; \citealp{RhPe98}) are characterised by radio shell with centre-filled X-rays. Recombining (overionized) plasmas (RPs) have been discovered from some MM SNRs (e.g. \citealp{Ya09, Sa12, Er14, Su18, Ka18}). Many of them have been also detected in gamma rays with energies ranging from GeV to TeV (e.g. \citealp{Ab10, Al07}). X-ray and gamma-ray studies of MM SNRs are important to address the questions about their relation with the interstellar medium (ISM), the physical origin of RPs (e.g. thermal conduction: \citealp{Co99, Sh99} and rarefaction: \citealp{It89, Sh12}) and the origin of the gamma-ray emission (e.g. leptonic and hadronic).

HB9 (G160.9+2.6) has a large angular size (140 arcmin $\times$ 120 arcmin) and a shell type morphology in the radio band (e.g. \citealt{LeRo91}).
The estimated distance and age by \citet{LeAs95} was found to be 1.5 kpc and 8$-$20 kyr from {\it ROSAT} data, respectively. Using H\,{\sc i} observations, the kinematic distance of HB9 was estimated to be 0.8 $\pm$ 0.4 kpc \citep{LeTi07}. They also estimated the age of HB9 as 4000$-$7000 yr.  

In X-rays, HB9 was first discovered with {\it HEAO-1} A2 soft X-ray survey in 1979  \citep{Tu79}, then observed by {\it GINGA} and {\it ROSAT} X-ray satellites \citep{Ya93, LeAs95}.
Using {\it GINGA} data, \citet{Ya93} found that the X-ray spectrum was represented by a two-component model: a thin thermal emission (0.4$-$0.7 keV) at a low-energy band and a thermal emission ($\sim$6.6 keV) or power-law (photon index of $\Gamma$ $\sim$ 2.35) models at a high energy band.
\citet{LeAs95} studied {\it ROSAT} data and presented the full {\it ROSAT} Position Sensitive Proportional
Counter (PSPC) image of HB9. From this image, they concluded that HB9 has a centrally brightened X-ray emission. They also searched for the spatial variation and found that the column density is nearly constant across HB9 and the electron temperature of the outer regions is slightly lower than the central region. The X-ray temperature shows no strong spatial variation \citep{LeAs95}. This result with the radio/X-ray morphology indicates that the SNR HB9 belongs to the class of MM SNRs. 

In the GeV gamma-ray band, \citet{Ar14} analysed  5.5 years of  data collected by the Large Area Telescope (LAT) on board the Fermi Gamma-ray Space Telescope ({\it Fermi}) and detected extended gamma-ray emission centered at the position of HB9 with a significance of 16$\sigma$ above 0.2 GeV. The radius of the disk-like extended gamma-ray source was found to be $\sim$1 degree and the spectrum was fit to a log-parabola function,  defined as $dN/dE=N_{0}(E/E_{0})^{-(\alpha+\beta~ln(E/E_{0}))}$, with spectral indices of $\alpha$ = 2.2 $\pm$ 0.09 and $\beta$ = 0.4 $\pm$ 0.1.  Although HB9 was reported to be `not detected' in the First {\it Fermi}-LAT Supernova Remnant Catalog \citep{Ac16}, it was listed as a new extended source, 4FGL J0500.3+4639e, in the Fourth {\it Fermi}-LAT sources (4FGL; \citealp{Fe19a}) catalog. 

The closest compact source to HB9 is the pulsar PSR B0458+46/PSR J0502+4654 that is located within the radio continuum shell of the SNR \citep{Kr17}. This pulsar is at a distance of 1.79 kpc and so far no gamma-ray emission has been detected from it.  The closest point-like gamma-ray sources, that are located outside the radio-shell of HB9 were reported in the 4FGL catalog and the Fourth (4LAC; \citealp{Fe19b}) catalog of active galactic nuclei (AGNe): 4FGL J0503.6+4518, a blazar (GB6 J0503+4517), and 4FGL J0507.9+4647, an AGN named as TXS 0503+466. 

In this paper, we explore the X-ray properties of the SNR HB9 using archival data of {\it Suzaku} \citep{Mit07}, which has a high spectral resolution and a low instrumental background.  In addition, we re-analyse the {\it Fermi}-LAT data of HB9 collected over the span of about 10 years to map its morphology in greater detail and investigate overlapping regions of gamma-ray emission with the ambient atomic and molecular gas, as well as the RP emission. The organization of the paper is as follows: In Section \ref{obs}, we describe the X-ray, gamma-ray, and atomic \& molecular gas observations and the data reduction processes. In Section \ref{Analyses and Results}, we outline the data analysis procedure and results. We then discuss our results in the context of other MM SNRs showing RP emission in Section \ref{discussion}. Finally, we present our conclusions in Section \ref{conclusions}. 

\section{Observations and data reduction}
\label{obs}
\subsection{X-rays}
 Two pointing observations (the east and west) of HB9 were performed in September of 2014 (PI: T. Pannuti) with X-ray imaging spectrometer (XIS; \citealt{Ko07}) on board {\it Suzaku}, as listed in Table 1. We downloaded archival {\it Suzaku} data\footnote{Data are available through the Data Archives and Transmission System (DARTS) at \url{https://darts.isas.jaxa.jp/astro/suzaku}} and extracted the XIS data from XIS 0, 1, 3. Note that XIS0 and XIS3 are front-side illuminated (FI) CCDs, whereas XIS1 is a back-side illuminated (BI) CCD. 
\begin{table*}
 \begin{minipage}{170mm}
 \begin{center}
  \caption{Log of X-ray observations of HB9 and background.} 
\begin{tabular}{@{}p{3.4cm}p{2.5cm}p{1.6cm}p{1.6cm}p{2.5cm}p{2.2cm}@{}}
  \hline
Target name        			& Observation ID 	          & Start time & Stop time & ($l, b$)       	&  Exposure time            \\ 
             			&                     &                & &  &     (ks)           \\
\hline
HB9  east  		        &     509033010          & 2014-09-29      & 2014-09-30    &  ($160\fdg8$, $2\fdg65$) & 51.1   	          \\ 

HB9  west		        &     509032010          & 2014-09-30    &  2014-10-01     & ($160\fdg5$, $2\fdg22$) &49.8  	          \\ 
IRAS 05262+4432 (BGD)		        &     703019010          & 2008-09-12    &  2008-09-14     & ($165\fdg1$, $5\fdg72$) & 82.1    	          \\ 
\hline
\label{Table1}
\end{tabular}
\end{center}
\end{minipage}
\end{table*}
In the reduction procedure and the analysis of the {\it Suzaku} data, we used {\sc headas} software version 6.20 and {\sc xspec} \citep{Ar96} version 12.9.1 with the latest atomic database AtomDB v3.0.9\footnote{\url{http://www.atomdb.org}} \citep{Sm01, Fo12}. We generated redistribution matrix files and ancillary response files with the ftool {\sc xisrmfgen} and {\sc xissimarfgen} \citep{Is07}, respectively. To extract XIS images and spectra we used {\sc xselect} v2.4d. The spectra were rebinned with a minimum of 25 counts to allow the use of the $\chi^2$-statistic.
\begin{figure*}
\centering \vspace*{1pt}
\includegraphics[width=0.60\textwidth]{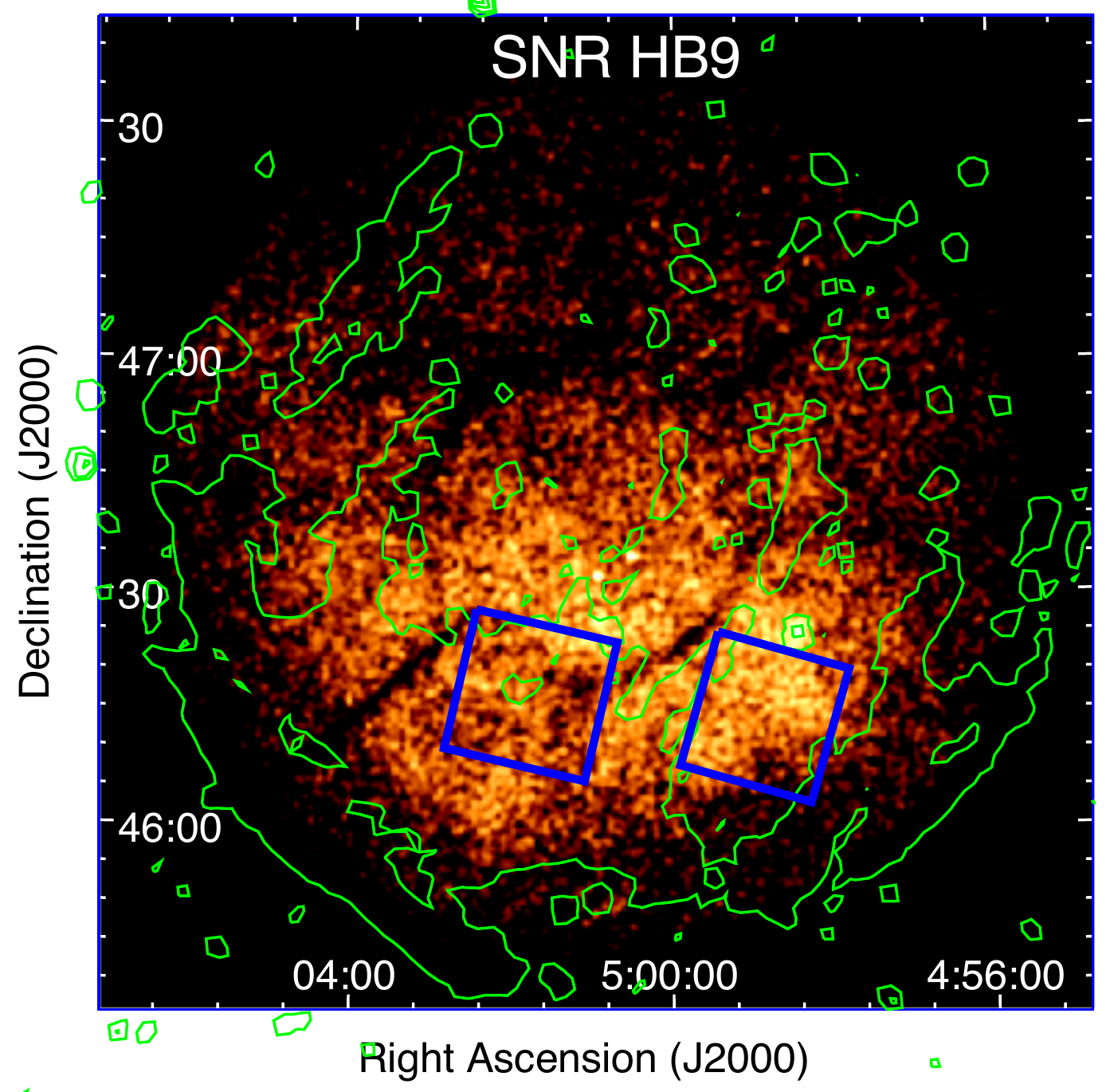}
\includegraphics[width=0.40\textwidth]{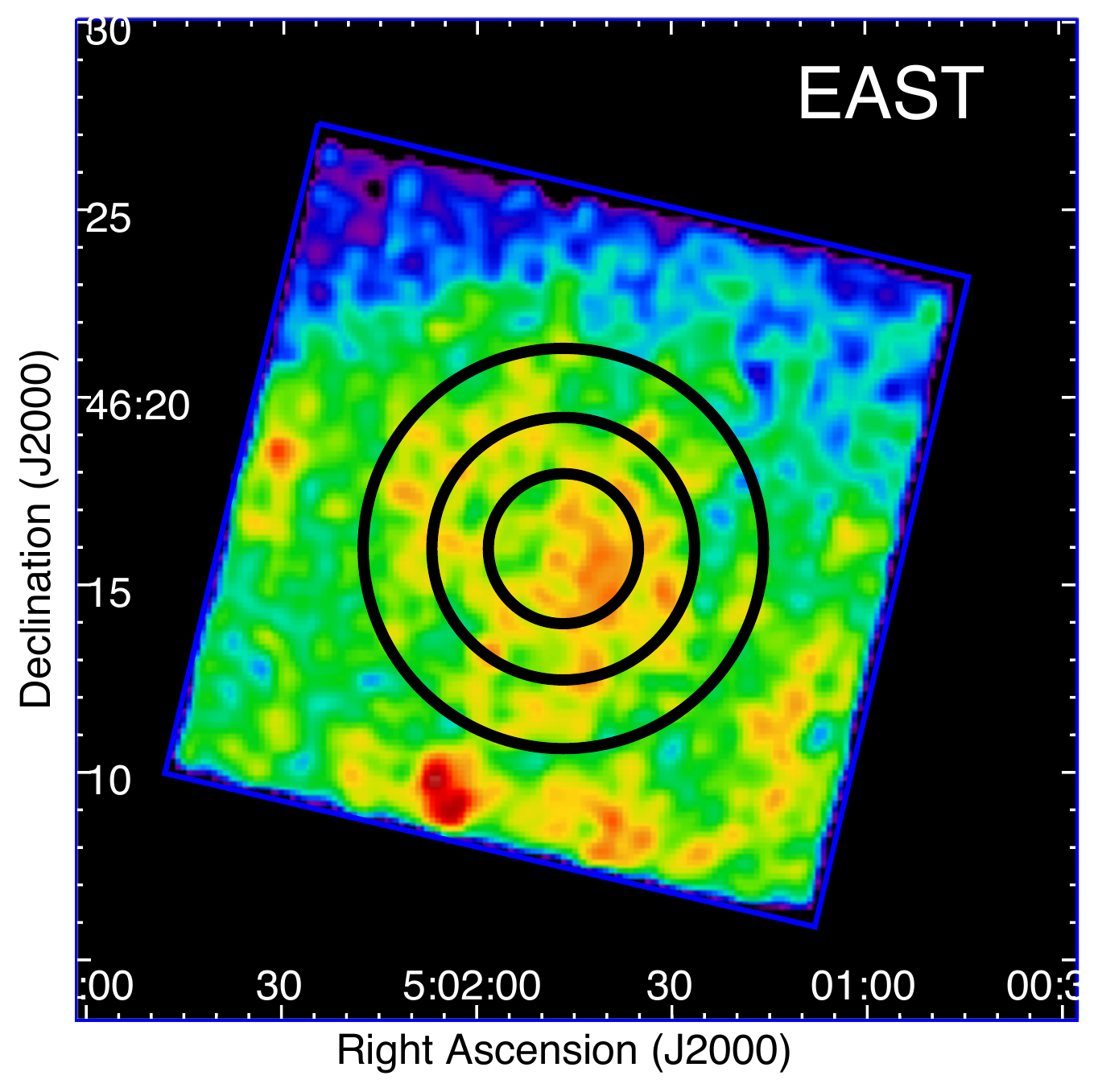}
\includegraphics[width=0.40\textwidth]{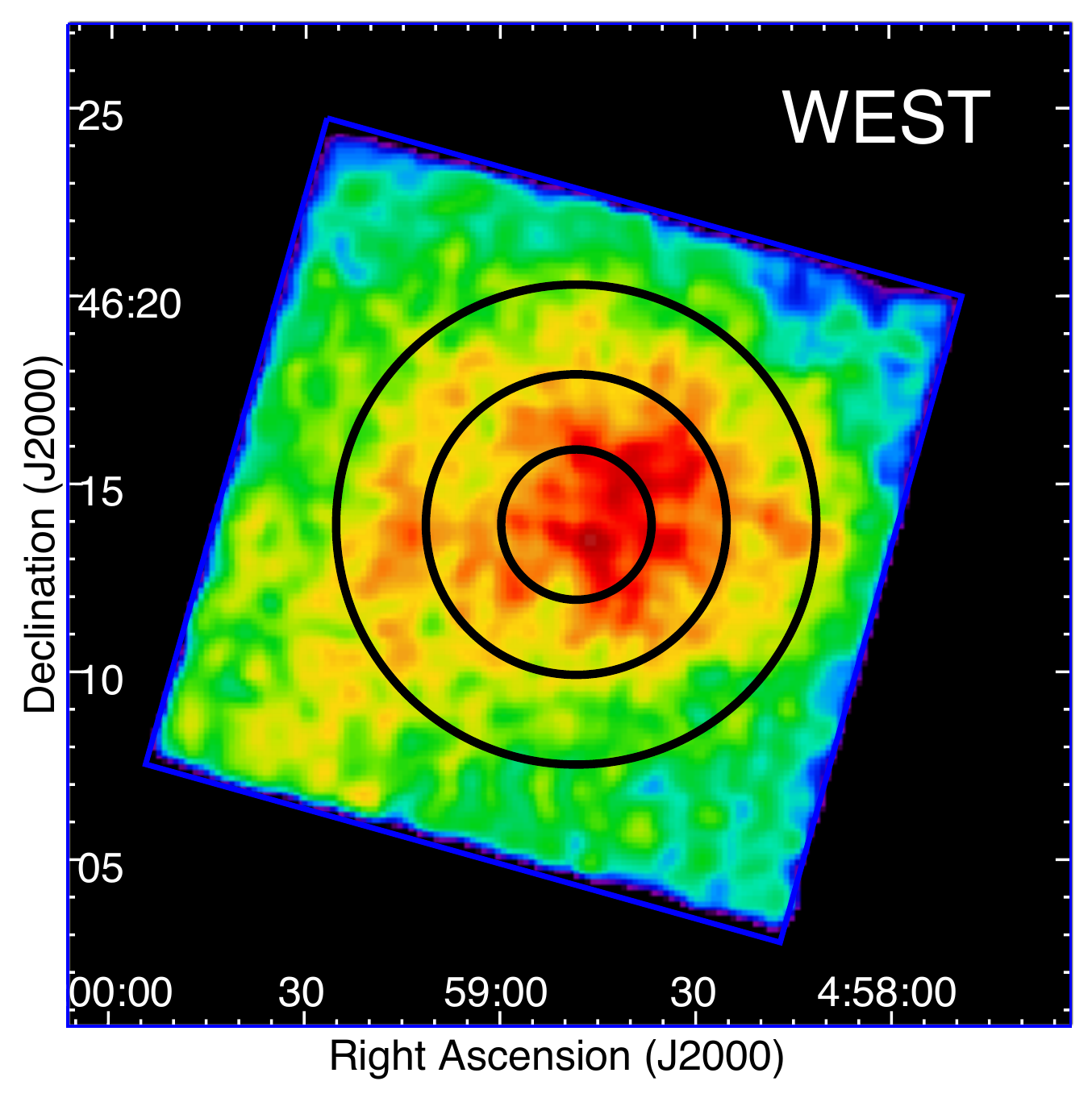}
\caption{Top: The {\it ROSAT} PSPC image of the HB9 field in the 0.2$-$2.4 keV band, overlaid with the radio continuum contours at 1420 MHz observed with the Synthesis Telescope (ST) of the Dominion Radio Astrophysical Observatory (DRAO). The radio contours are at levels of 0.5, 1.6, 5.8,
12.1, 25.7 and 54.9 mJy beam$^{-1}$. The individual squares indicate the {\it Suzaku} XIS field of views (FoVs) for the east and west observations given in Table 1. Bottom left: XIS1 image of the east region of HB9 in the 0.3$-$10.0 keV band. Bottom right: XIS1 image of the west region of HB9 in the 0.3$-$10.0 keV band.  The circles in the bottom panels illustrate regions used for spectral analysis. North is up and east is to the left.}
\label{Fig1}
\end{figure*}

\subsection{Gamma rays}
The gamma-ray observations were taken from 2008-08-04 to 2018-11-14.  In this analysis,  we made use of the analysis packages \texttt{fermitools}\footnote{\url{http://fermi.gsfc.nasa.gov/ssc/data/analysis/software}} version \texttt{1.0.1} and  \texttt{fermipy}\footnote{\url{http://fermipy.readthedocs.io/en/latest/index.html}} version \texttt{0.17.4}. Using \texttt{gtselect} of \texttt{fermitools} we selected {\it Fermi}-LAT Pass 8 `source' class and `front$+$back' type events coming from zenith angles smaller than 90$^{\circ}$ and from a circular region of interest (ROI) with a radius of 20$^{\circ}$ centered at the SNR's radio location\footnote{SNR's radio location at  R.A.(J2000) = 75$^{\circ}\!\!$.25 and decl.(J2000) = 46$^{\circ}\!\!$.67.}. The {\it Fermi}-LAT instrument response function version \emph{P8R3$_{-}$SOURCE$_{-}\!\!$V2} was used. For mapping the morphology and searching for new sources within the analysis region, events having energies in the range of 1$-$300 GeV were selected. To deduce the spectral parameters of the investigated sources, events with energies between 200 MeV and 300 GeV were chosen.

\subsection{CO, H\,{\sc i} and radio continuum}
We used archival $^{12}$CO($J$ = 1--0) data sets obtained with the 1.2 m telescope at Harvard-Smithsonian Center for Astrophysics (CfA) using the position-switching technique \citep{Da01}. The angular resolution is $\sim$8.8 arcmin. The sampling interval of the HB9 region is $0\fdg0625$ for $|b|$ $<$ 2$^{\circ}$ and $0\fdg25$ for $|b|$ $>$ 2$^{\circ}$. The typical noise fluctuation is $\sim$0.05 K ($|b|$ $<$ 2$^{\circ}$) or $\sim$0.14 K ($|b|$ $>$ 2$^{\circ}$) at the velocity resolution of $\sim$1.3 km s$^{-1}$.

The H\,{\sc i} 21 cm line and radio continuum at 408 MHz and 1420 MHz are from the Canadian Galactic Plane Survey (CGPS; \citealt{Ta03}), which were carried out at the Dominion Radio Astrophysical Observatory (DRAO). These datasets have been also published by \citet{LeTi07}. The angular resolution is 58 arcsec $\times$ 80 arcsec for H\,{\sc i}; 2.8 arcmin $\times$ 3.9 arcmin for 408 MHz; 49 arcsec $\times$ 68 arcsec for 1420 MHz. The typical noise fluctuation of H\,{\sc i} is $\sim$3 K at the velocity resolution of $\sim$0.82 km s$^{-1}$. The radio continuum images are noise limited with root-mean-square of $\sim$3 mJy beam$^{-1}$ at 408 MHz and $\sim$0.3 mJy beam$^{-1}$ at 1420 MHz.

\section{Analysis and Results}
\label{Analyses and Results}
\subsection{{\it Suzaku} analysis \& results}

\subsubsection{X-ray image}
We display a {\it ROSAT} PSPC image of HB9 in the 0.2$-$2.4 keV energy band in the top panel of Fig. \ref{Fig1}. The contours correspond to radio observations\footnote{The data downloaded here:\url{http://www.cadc-ccda.hia-iha.nrc-cnrc.gc.ca/en/search}} at 1420 MHz with the Synthesis Telescope (ST) of the DRAO. We also show 0.3$-$10.0 keV XIS1 images of the east and west regions of HB9 in the left and right bottom panels of Fig. \ref{Fig1}, respectively. To investigate the plasma parameters, we selected regions which are shown by the circles on these images.

\subsubsection{Background estimation}
We estimated the background spectra from the observational {\it Suzaku} data, IRAS 05262+4432. The observation log for the background (BGD) is presented in Table 1. For the instrumental background (NXB: non X-ray background), we used {\sc xisnxbgen} ftool \citep{Ta08} and subtract it from the extracted spectrum. The X-ray background spectral model in our analysis includes four components: the Cosmic X-ray background (CXB), the local hot bubble (LHB) and two thermal components (LP: low temperature plasma and HP: high temperature plasma) for the Galactic halo (GH). Similar to \citet{Ya93}, the contribution of the Galactic ridge X-ray emission (GRXE) is negligible in our analysis, since it is located near the anti-center region. 

We fit the spectra with this model: Abs1 $\times$ power-law$_{\rm CXB}$ + Abs2 $\times$ (apec$_{\rm HP}$ + apec$_{\rm LP}$) + Abs3 $\times$ apec$_{\rm LHB}$. Abs1, 2, 3 represent the ISM absorption for CXB, GH and LHB, respectively. The apec is a collisional ionisation equilibrium (CIE) plasma model in {\sc xspec}. We left the normalization parameters of each component free and fixed the electron temperature parameters of all components at the values were set by \citet{Ma09}. The CXB component parameters are fixed to the values given by \citet{Ku02}. 

\subsubsection{Spectral fitting}
In order to explore the X-ray spectral properties of HB9, we extracted XIS spectra from the east and west regions indicated with the outermost circles (radii of 5.3 and 6.4 arcmin) shown in Fig. \ref{Fig1}. We see emission lines of highly ionized O, Ne, Mg, Si and Fe-L below 2 keV.

\begin{figure*}
\includegraphics[width=0.43\textwidth]{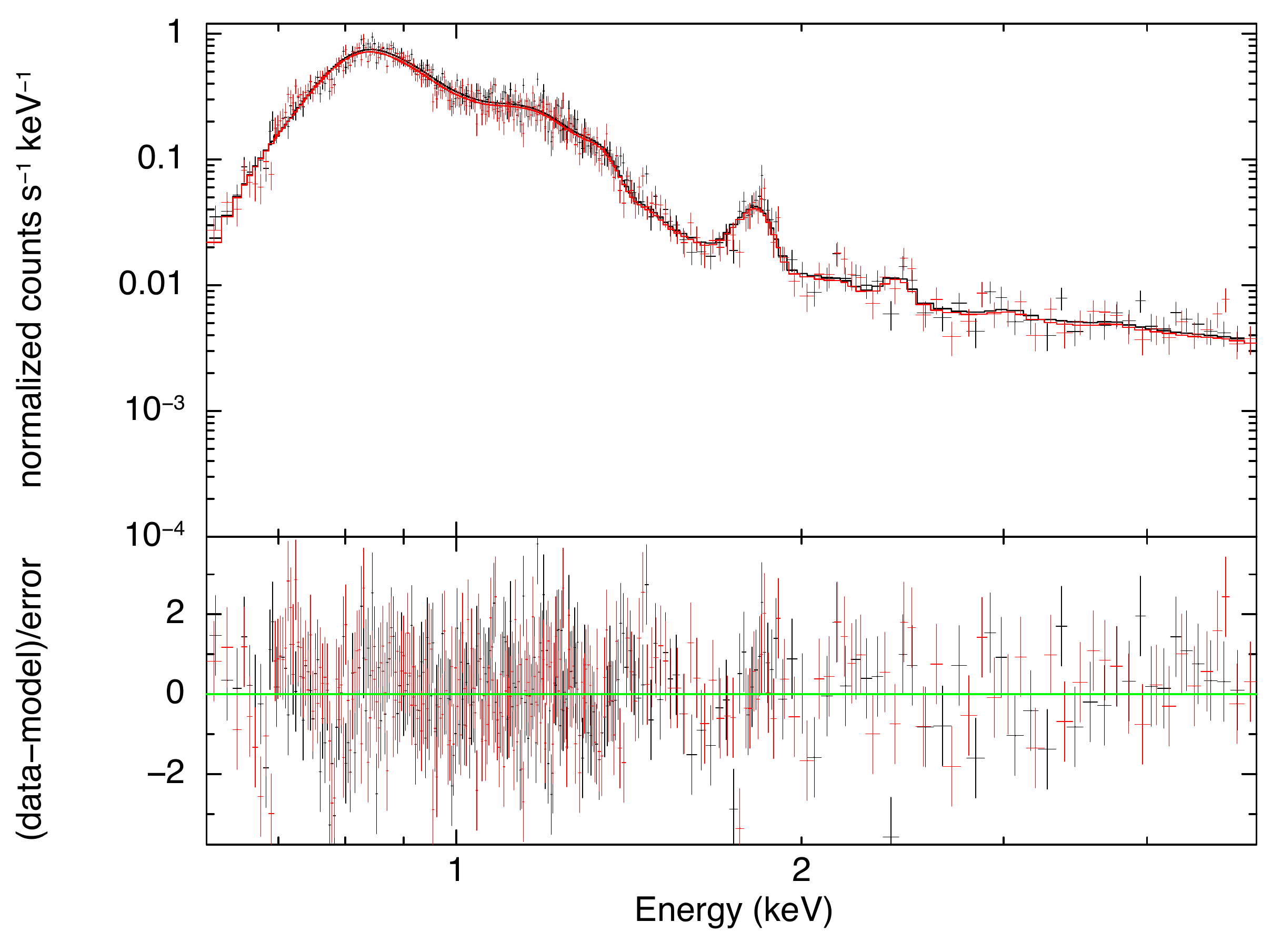}
\includegraphics[width=0.43\textwidth]{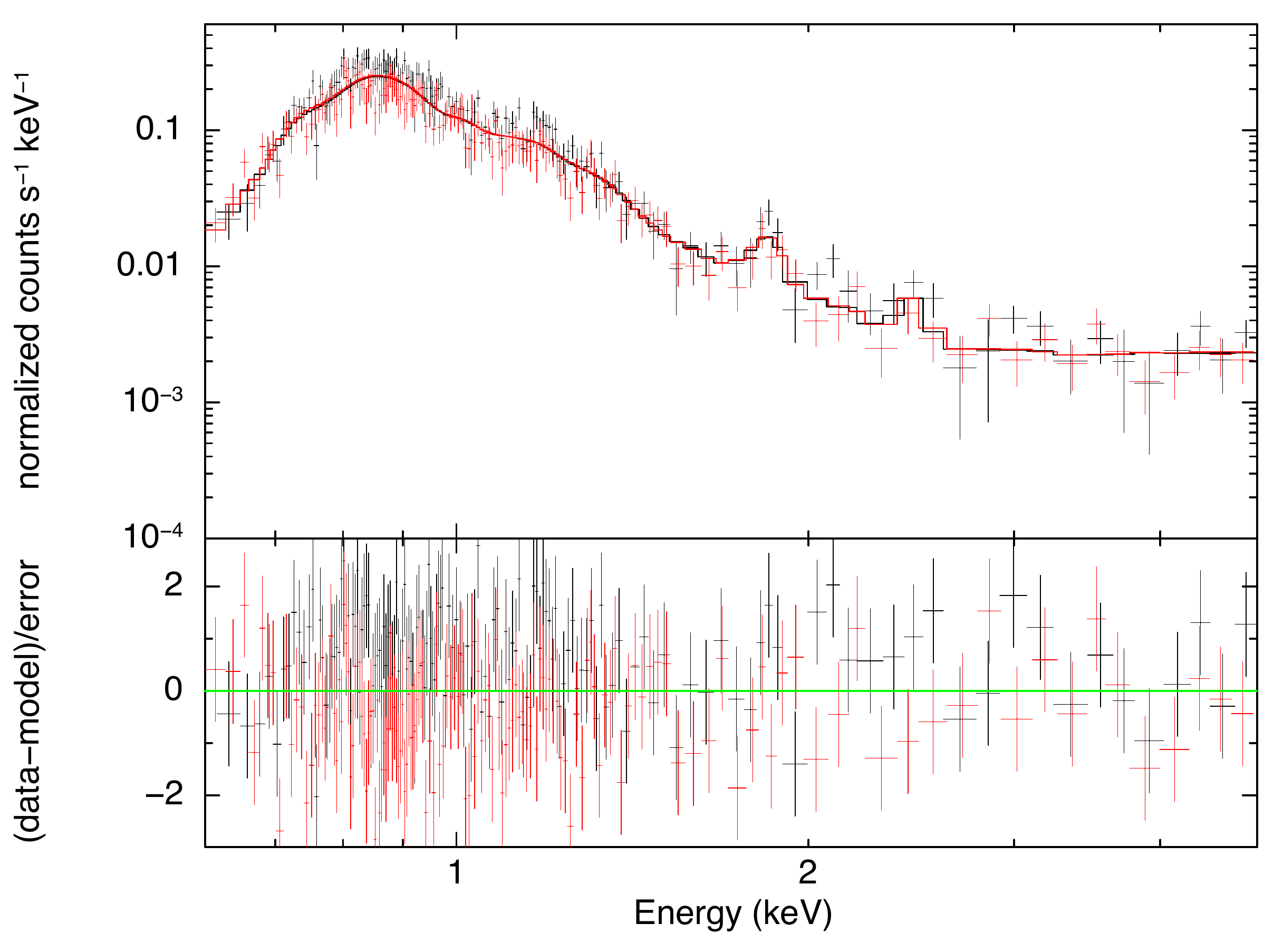}
\caption{Left: {\it Suzaku} XIS FI spectra of the west region in the 0.6$-$5.0 keV energy band. Right: XIS FI spectra of the east region in the 0.6$-$5.0 keV energy band. The residuals are shown in the lower panel.}
\label{Fig2}
\end{figure*}

\begin{table*}
\begin{minipage}{170mm}
\begin{center}
\caption{{\it Suzaku} best-fitting spectral parameters for the west and east {\it Suzaku} observation regions of HB9, shown in Fig. \ref{Fig1}.}
\renewcommand{\arraystretch}{1.5}
\begin{tabular}{@{}p{2.9cm}p{3.9cm}p{2.9cm}p{2.5cm}p{2.5cm}@{}}
\hline\hline
 &&\multicolumn{2}{c}{Value}\\
    \cline{3-4}
Component & Parameter (Unit)                            & West                        & East                                     \\
\hline
TBABS     &	$N_{\rm H}$ ($10^{21}$ cm$^{-2})$            &    $4.76_{-0.21}^{+0.44}$       &       $3.57_{-0.33}^{+0.42}$       \\
VRNEI    &	  $kT_{\rm e}$ (keV)                         &    $1.13_{-0.03}^{+0.02}$       &       $0.97_{-0.02}^{+0.01}$       \\
         &	  $kT_{\rm init}$ (keV)                      &    $3.14_{-0.25}^{+0.41}$       &       0.0808 (fixed)               \\
         &	 O (solar)                                   &    $1.2_{-0.3}^{+0.2}$          &       $1.1_{-0.4}^{+0.2}$          \\
         &	 Ne (solar)                                   &    $1.1_{-0.1}^{+0.2}$          &       $1.3_{-0.2}^{+0.1}$          \\
         &	 Mg (solar)                                  &    $1.3_{-0.2}^{+0.1}$          &       $1.2_{-0.4}^{+0.2}$          \\
         &	 Si (solar)                                  &    $1.7_{-0.1}^{+0.2}$          &       $1.5_{-0.3}^{+0.2}$          \\
         & $\tau=n_{\rm e}t$ ($10^{11}$ cm$^{-3}$ s)                &    $5.8_{-0.4}^{+0.3}$          &       $3.6_{-0.5}^{+0.2}$        \\
         & Normalization$\dagger$  ($10^{-3}$ cm$^{-5}$)          &    $2.25_{-0.12}^{+0.17}$       &       $3.21_{-0.26}^{+0.41}$       \\

CIE     &	 $kT_{\rm e}$ (keV)                          &     $0.42_{-0.02}^{+0.01}$      &       $0.51_{-0.01}^{+0.02}$         \\
         &	 Normalization$\dagger$   ($10^{-3}$ cm$^{-5}$)        &     $2.81_{-0.27}^{+0.21}$      &       $4.13_{-0.51}^{+0.25}$         \\
\hline
         &	 Reduced $\chi^{2}$ (dof)                    &    1.09 (587)                   &      1.11 (498)                        \\                
 \hline
\label{Table2}
\end{tabular}
\end{center}
\begin{tablenotes}
\item {\bf Notes.} Errors are within a 90 per cent confidence level. Abundances are given relative to the solar values of \citet{Wi00}. \\
\item $\dagger$ The unit is $10^{-14}$ $\int n_{\rm e} n_{\rm H} dV$/($4\pi d^{2}$), where $d$ is the distance to the source (in cm), $n_{\rm e}$ and $n_{\rm H}$ are the electron and hydrogen densities (in units of cm$^{-3}$), respectively, and $V$ is the emitting volume (in units of cm$^{3}$). \\
\end{tablenotes}
\end{minipage}
\end{table*}

We first examined XIS spectra of the west region by fitting with various single component models that include non-equilibrium ionization (NEI), vnei, plane-parallel shock model, vpshock, CIE, vmekal and vapec, RP model, VRNEI in {\sc xspec}. Each model modified by the interstellar absorption model TBABS \citep{Wi00}. They failed to account for the emission above 3.0 keV and did not give adequately good fits (reduced $\chi^{2}$$>$1.8). 

We subsequently fitted the spectra with a two-component model, NEI and CIE model, to better reproduce the spectra. However, this model did not provide a good fit to the data (reduced $\chi^{2}$= 1.5$-$1.8).  We then applied RP and CIE model. The RP model has variable abundances (VRNEI in {\sc xspec}), it corresponds to the spectrum of a NEI plasma after a rapid transition of the electron temperature from $kT_{\rm init}$ to $kT_{\rm e}$. In this fitting, the free parameters are the current electron temperature ($kT_{\rm e}$), initial temperature ($kT_{\rm init}$), ionization time-scale ($n_{\rm e}t$), normalization, and abundances of O, Ne, Mg and Si relative to the solar values of \citet{Wi00}. For CIE component, the electron temperature and normalization are free parameters. The elemental abundances in the CIE component are fixed at the solar values. We also added narrow lines (Gaussian models) at 0.82 and 1.23 keV for the Fe-L lines. The two-component model, an RP and a CIE model, improve the fitting with reduced $\chi^{2}$/dof=1.09. We found that an initial temperature ($kT_{\rm init}$ $\sim$ 3.14 keV) larger than the final temperature ($kT_{\rm e}$ $\sim$ 1.13 keV) indicates that the plasma is in the RP state. The fitting results are given in Table \ref{Table2}. The spectra extracted from the west region are shown in Fig. \ref{Fig2}.

For the east region, we constrained above models and found that the spectra are best reproduced by the model consisting of the NEI and CIE components with reduced $\chi^{2}$/dof=1.11. The free parameters were the absorption column density ($N_{\rm H}$), electron temperature ($kT_{\rm e}$), ionization time-scale ($n_{\rm e}t$), normalisation and elemental abundances of O, Ne, Mg and Si. Those of the other elements are fixed to the solar values. We fixed $kT_{\rm init}$ to 0.0808 keV, which indicates that the plasma is in NEI condition. The all elements in the CIE component are fixed at the solar values. The NEI model requiring a low ionization time-scale and slightly enhanced abundance for Si. Fig. \ref{Fig2} shows the XIS spectra of the east region. The best-fitting parameters are reported in Table \ref{Table2}.

Next we examine the possible radial variations in the X-ray spectral parameters by extracting the XIS spectra from the three concentric annulus regions both for the west and east regions (see the bottom panels of Fig. \ref{Fig1}). Our analysis confirm the results of  \citet{LeAs95} that there are no significant variation of parameters, temperature, ionization time-scale ($\tau$) and metal abundances. In this fitting, we kept the $N_{\rm H}$ values at its best-fitting value for the whole region.

\subsection{Gamma-ray analysis \& results}
\subsubsection{The background model}
In the background model, the galactic (\emph{gll$_{-}$iem$_{-}$v7.fits}) and the isotropic (\emph{iso$_{-}$P8R3$_{-}$SOURCE$_{-}\!\!$V2$_{-}\!\!$v1.txt}) diffusion components, as well as all the sources from 4FGL catalog within the ROI were included. 

\begin{figure*}
\centering \vspace*{1pt}
\includegraphics[width=0.9\textwidth]{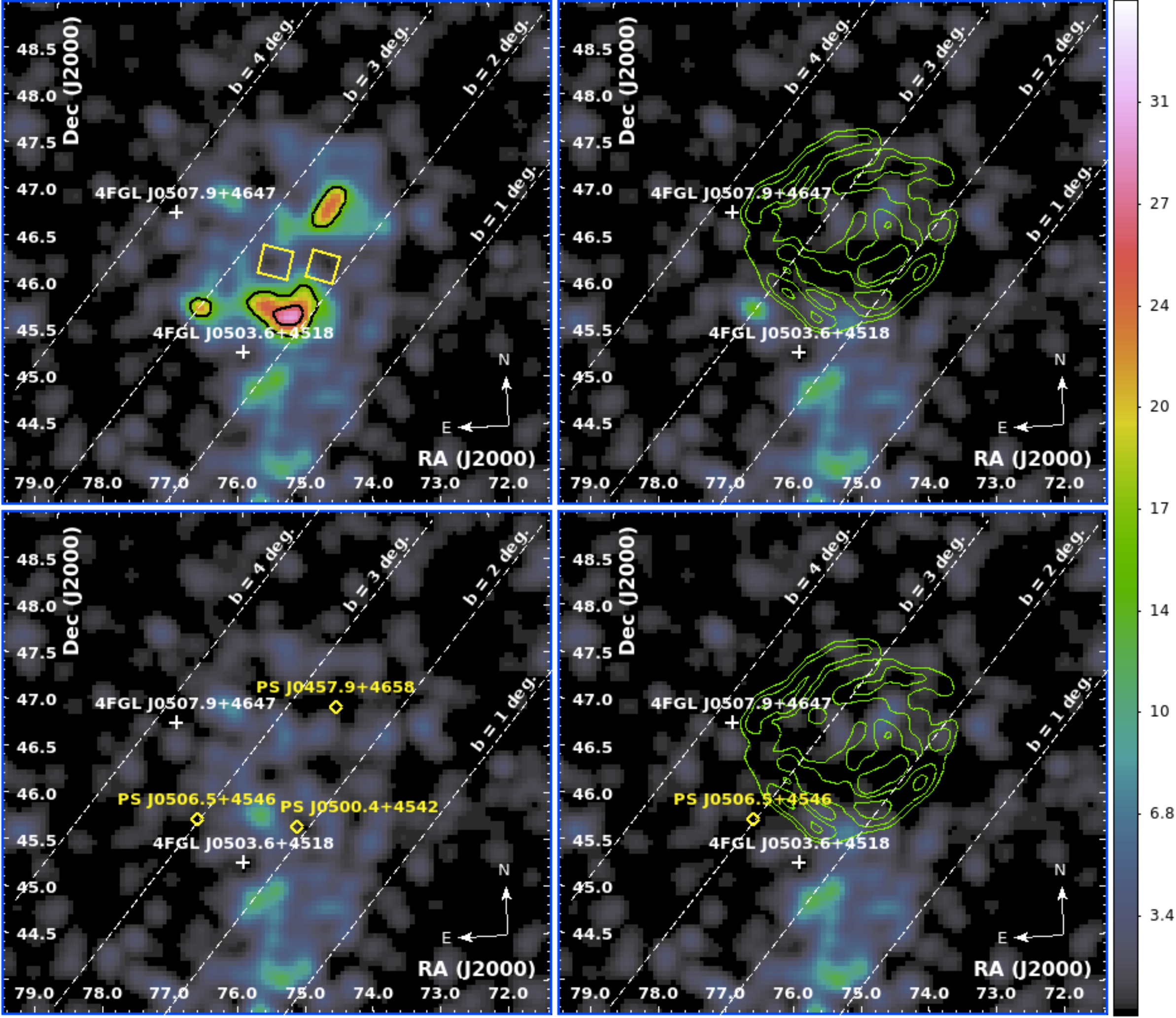}
\caption{The gamma-ray TS maps in the energy range of 1 - 300 GeV. All maps given in the R.A. - Decl. (J2000) coordinate system in radio continuum with a bin size of 0.05$^{\circ}\times~$0.05$^{\circ}$. The white plus markers with bold-written labels show the point sources from the 4FGL catalog. The white dashed lines show the latitude levels in the Galactic coordinates. {\bf Top-left Panel:} The map is produced without the inclusion of HB9 into the gamma-ray background model.  The black contours represent the significance levels of gamma rays at 4$\sigma$ and 5$\sigma$. The two big yellow squares are the {\it Suzaku} regions analysed in this paper. {\bf Top-right Panel:} The TS map produced including the Radio Template model of HB9 (see Sect. 3.2.2) into the gamma-ray background model. Green contours show radio data taken by the Green Bank telescope at 4850 MHz (600, 1095, 2190, 3285, 4380 Jy beam$^{-1}$). {\bf Bottom-left Panel:} The gamma-ray background model of this TS map includes three candidate point-sources, PS J0457.9+4658, PS J0500.4+4542, and PS J0506.5+4546, shown with yellow diamond markers. Here, the gamma-ray background model does not include HB9 as a gamma-ray source. {\bf Bottom-right Panel:} TS map, where the Radio Template model of HB9 and PS J0506.5+4546 (shown with a yellow diamond marker) are added into the gamma-ray background model. Green contours are the same as in the top-right panel.} 
\label{Fig3}
\end{figure*}

With \texttt{gtlike}, we performed the maximum likelihood \citep{Ma96} fitting on data that was binned within the selected energy and spatial ranges. During the fit, the normalization parameters of all the sources within 3$^{\circ}$ ROI, as well as the diffuse emission components were left free. Specifically, we freed the normalization parameter of all sources with significance\footnote{The detection significance value is approximately equal to the square root of the test statistics (TS) value. Larger TS values indicate that the null hypothesis (maximum likelihood value for a model without an additional source) is incorrect.} $>$ 20 and we fixed all parameters for sources with significance $<$ 20. The parameters of all the other sources were fixed to their 4FGL-catalog-mentioned values.  

\subsubsection{HB9 source morphology}
Fig. \ref{Fig3} top-left panel shows the initial TS map produced in the energy range of 1$-$300 GeV for the 10$^{\circ}$ $\times$ 10$^{\circ}$ analysis region, where the gamma-ray background model did not contain HB9. In this map, three relatively bright peaks (significance higher than 4$\sigma$) are visible. However, none of these bright peaks coincide with the {\it Suzaku} observation regions depicted with the two yellow squares on Fig. \ref{Fig3} top-left panel. 
\begin{table}
    \caption{Spatial fit results of gamma-ray data for HB9 produced in the energy range of 1$-$300 GeV.  PL and LP stands for power-law and log-parabola type spectrum, respectively (see Section 3.2.4 for definitions of PL and LP).}
    \begin{minipage}{0.45\textwidth}
    	\centering
	\begin{tabular}{lccc} 
        \hline\hline
        Spatial Model                     &Spectral Model &$TS_{\rm ext}$  \\
 		\hline
         \\
         \multirow{2}{*}{Radial Disk}     &LP             &78 \\
                                          &PL             &83 \\
         \\
         \multirow{2}{*}{Radial Gaussian} &LP             &101 \\
                                          &PL             &104 \\
          \\                     
         \multirow{2}{*}{Radio Template}  &LP             &109   \\
                                          &PL             &109   \\
        \hline 
        \label{table_3}
	\end{tabular}
	      \vspace{-0.3cm}
    \end{minipage}
\end{table}
Using the \texttt{extension} method of the \texttt{fermipy} analysis package, we fit three different types of spatial models to the excess gamma-ray distribution found between 1 and 300 GeV: A Radial Disk model, Radial Gaussian model, as well as the spatial template of the 4850 MHz radio continuum data from the Green Bank telescope \citep{Co94} (herewith Radio Template model).
The \texttt{extension} method computes a likelihood ratio test with respect to the point-source hypothesis and applies a best fit model for the extension. The best-fitted extension is found by performing a likelihood profile scan over the source width.
\par To find the significance of the extension measurement, we used the TS of the extension (TS$_{\rm ext}$) parameter, which is the likelihood ratio comparing the likelihood for being a point-like source ($L_{\rm pt}$) to a likelihood for an existing extension ($L_{\rm ext}$), TS$_{\rm ext}$ = -2log($L_{\rm ext}$/$L_{\rm pt}$). Table \ref{table_3} shows the fit results of these extension models where each spatial model is tested with both  power-law (PL) and a log-parabola (LP) type spectrum. Comparing the TS$_{\rm ext}$ values, we found that the Radio Template model is the best fitting extension model to the data with a TS$_{\rm ext}$ value of 109. 

\subsubsection{Testing for multiple gamma-ray point sources}
We re-produced the TS map including the Radio Template model with an LP-type spectrum into the gamma-ray background model. In the TS map shown in Fig. \ref{Fig3} top-right panel, we found no excess gamma-ray emission from the direction of the pulsar PSR B0458+46/PSR J0502+4654. However, some gamma-ray excess remains to be visible outside the southeast region of the SNR's radio shell as shown in Fig. \ref{Fig3} top-right panel.

The excess gamma-ray emission seen on Fig. \ref{Fig3} top-left panel could be a result of three point-like gamma-ray sources rather than a single extended source. So, we used the iterative source-finding algorithm \texttt{find$_{-}$sources()} in \texttt{fermipy} to search for the exact TS value and location of these point-source candidates on the TS map. This algorithm takes the peak detection on a TS map to find new source candidates. The algorithm identifies peaks with a significance threshold value higher than 5$\sigma$ and taking an angular distance of at least 1$^{\circ}\!\!$.5 from a higher amplitude peak in the map. It orders the peaks by their TS values and adds a source at each peak starting from the highest TS peak. Then it sets the source position by fitting a 2D parabola to the log-likelihood surface around the peak maximum. After adding each 5$\sigma$ source, it re-fits the spectral parameters of that source. With this algorithm we identified three new candidate point sources (PS J0457.9+4658, PS J0500.4+4542, PS J0506.5+4546) within the analysis region, two of which are located inside the Radio Template model of HB9. One source is located outside the SNR radio shell of HB9 (see Table \ref{table_4}). By adding all of these candidate point-like sources into the gamma-ray background model and excluding the best-fit extended model of HB9, the TS map in the bottom-right panel of Fig. \ref{Fig3} is obtained, which still shows some low-level gamma-ray excess within the radio shell of HB9.

To be able to compare the Radio Template model of HB9 with the model that contains only the three candidate point-like sources (herewith 3 Point Sources model), as well as other models, such as the Radio Template model plus a point-like source, PS J0506.5+4546 (herewith Radio Template + 1P.S. model), we implemented the Akaike Information Criterion (AIC) \citep{Ak98,La12} for each model. AIC is given by the following equation: 
\begin{equation}
 \\
 AIC=2k-2ln(L), 
\end{equation}
where $k$ is the number of estimated parameters in the model and $L$ is the maximum value of the likelihood function for the model. The best source model is considered to be the one that minimizes the AIC value. So, $\Delta$AIC = (AIC)$_{1}$ $-$ (AIC)$_{m}$ is used to compare the model yielding the maximum AIC value, the 3 Point Sources model with k=6 (given by the index value 1), with other models tested in this analysis. The other tested models are: Radio Template model (k=3, m=2), Radio Template + 1P.S. model (k=5, m=3), Radio Template + 2P.S. model (k=7, m=4) and Radio Template + 3P.S. model (k=9, m=5). The Radio Template + 2P.S. model contains the gamma-ray point-source candidates PS J0506.5+4546 and PS J0457.9+4658, as well as the best-fit extended model (the Radio Template model). The Radio Template + 3P.S. model contains the point-source candidates PS J0506.5+4546, PS J0500.4+4542, PS J0457.9+4658 and the Radio Template model. Both Radio Template + 2P.S. and Radio Template + 3P.S. models show significant improvement in describing the extended morphology of HB9 in terms of AIC. However, in both models the significance of the candidate point sources (PS J0457.9+4658 and PS J0500.4+4542) drop below 5$\sigma$. Therefore, these models were not considered further in the analysis. As a result, the Radio Template + 1P.S. model is the next best-fit model in describing the extended morphology of HB9 (see Table \ref{table_5}), where the Radio Template model represents the SNR HB9 (total TS$\sim$616) and PS J0506.5+4546 is representing a point-like source outside the radio shell of HB9 (total TS$\sim$30). When TS map is computed using this best-fit model inside the gamma-ray background model, we see no more excess gamma-ray emission from within the SNR (see bottom-right panel of Fig. \ref{Fig3}).
\begin{table}
    \caption{Locations of the three gamma-ray point-source candidates calculated in the energy range of 1 $-$ 300 GeV using the 3 Point Sources model. For all fits only statistical errors are given.}
    \begin{minipage}{0.45\textwidth}
    	\centering
	\begin{tabular}{l ccc} 
        \hline\hline
       Point-source                         &R.A.                &Decl.     \\
        Name                                &(J2000)             &(J2000) \\
 		\hline
         \\                     
          \multirow{1}{*}{PS J0457.9+4658}  &74$^{\circ}\!\!$.49 &46$^{\circ}\!\!$.97 \\
          \\                     
          \multirow{1}{*}{PS J0500.4+4542}  &75$^{\circ}\!\!$.11 &45$^{\circ}\!\!$.70 \\
          \\                     
          \multirow{1}{*}{PS J0506.5+4546}  &76$^{\circ}\!\!$.64 &45$^{\circ}\!\!$.78 \\
        \hline 
        \label{table_4}
        \vspace{-0.2cm}
	\end{tabular}
    \end{minipage}
\end{table}
\begin{table}
    \caption{Spatial fit results for the gamma-ray morphology of HB9. In column (1), 3 Point Sources model: The three point-source candidates are PS J0457.9+4658, PS J0500.4+4542, PS J0506.5+4546, all with PL-type spectra; Radio Template+1P.S. model: Radio spatial template is modelled together with PS J0506.5+4546, which has a PL-type spectrum. (2)nd column shows the degrees of freedom (d.o.f.) of each model and the (3)rd column gives $\Delta$AIC value. }
    \begin{minipage}{0.45\textwidth}
    	\centering
	\begin{tabular}{lccc} 
        \hline\hline
        Spatial Model                          &d.o.f. &$\Delta$AIC   \\
        $~~~~~~$(1)                            &(2)    &(3)  \\ 
 		\hline
           \\
         \multirow{1}{*}{3 Point Sources}      &6      &0   \\
         \\
         \multirow{1}{*}{Radio Template}       &3      &11\\
         \\                     
         \multirow{1}{*}{Radio Template+1P.S.} &5      &41    \\
        \hline 
        \label{table_5}
        \vspace{-0.3cm}
	\end{tabular}
    \end{minipage}
\end{table}

\subsubsection{Gamma-ray spectrum}
\label{section324}
 During the extension measurements of HB9 (see Section 3.2.2), we used the following spectral models:
 \[
\mbox{- For PL:}~~~~~~~~~~\mbox{dN/dE} = \mbox{N}_0~\mbox{E}^{-\Gamma}
  \]
  \vspace{-0.8cm}
 \[
\mbox{- For LP:}~~~~~~~~~~\mbox{dN/dE} =  \mbox{N}_0 ~(\mbox{E}/\mbox{E}_b)^{{\rm-(\alpha + \beta
 \log(E/E_{b}))}}
 \]
where E$_b$ is a scale parameter. $\Gamma$ and (${\rm \beta}$, ${\rm \alpha}$) are spectral indices of the PL and LP spectral models, respectively. N$_0$ is the normalization parameter. We also applied the PL model during the point-sources search in Section 3.2.3. These models were used again during the spectral measurements performed in the 200 MeV $-$ 300 GeV energy range. 

The TS value of HB9 was found to be higher using the LP-type spectrum (TS = 627) in comparison to the PL-type spectrum (TS = 598). So, we continued the analysis using the LP-type spectral model for HB9.

Fitting the Radio Template+1P.S. model within the energy range of 200 MeV $-$ 300 GeV resulted in the following spectral parameters for both sources: 
\begin{itemize}
    \item {\it HB9 described by the Radio Template model}: The LP-spectral indices were found to be ${\rm \alpha}$ = 2.36 $\pm$ 0.05 and ${\rm \beta}$ = 0.14 $\pm$ 0.05. The total flux and energy flux values were found to be (2.47 $\pm$ 0.12)$\times$10$^{-8}$ cm$^{-2}$ s$^{-1}$ and (1.51 $\pm$ 0.08)$\times$10$^{-5}$ MeV cm$^{-2}$ s$^{-1}$ for $E_{b}$ parameter fixed to 747 MeV. 
    
    \item {\it PS J0506.5+4546 described by a point-source model}: The PL-spectral index was found to be ${\rm \Gamma}$ = -1.90 $\pm$ 0.19. The total flux and energy flux values were found to be (6.59 $\pm$ 3.47)$\times$10$^{-10}$ cm$^{-2}$ s$^{-1}$ and (1.29 $\pm$ 0.38)$\times$10$^{-6}$ MeV cm$^{-2}$ s$^{-1}$.
\end{itemize}

For HB9, the spectral parameter $\alpha$ was found to be consistent with the one reported by \citet{Ar14} in the same energy range and it agreed with result given in the 4FGL source catalog \citep{Fe19a}. However, the $\beta$ spectral parameter is found to be lower than the one ($\beta$ = 0.4 $\pm$ 0.1) found by both \citet{Ar14} and \citep{Fe19a}.

\vspace{-0.5cm}
\subsection{CO and H\,{\sc i} analysis results}
\label{CO_HI_Data_Analysis}
\begin{figure*}
\centering \vspace*{1pt}
\includegraphics[width=0.98\textwidth]{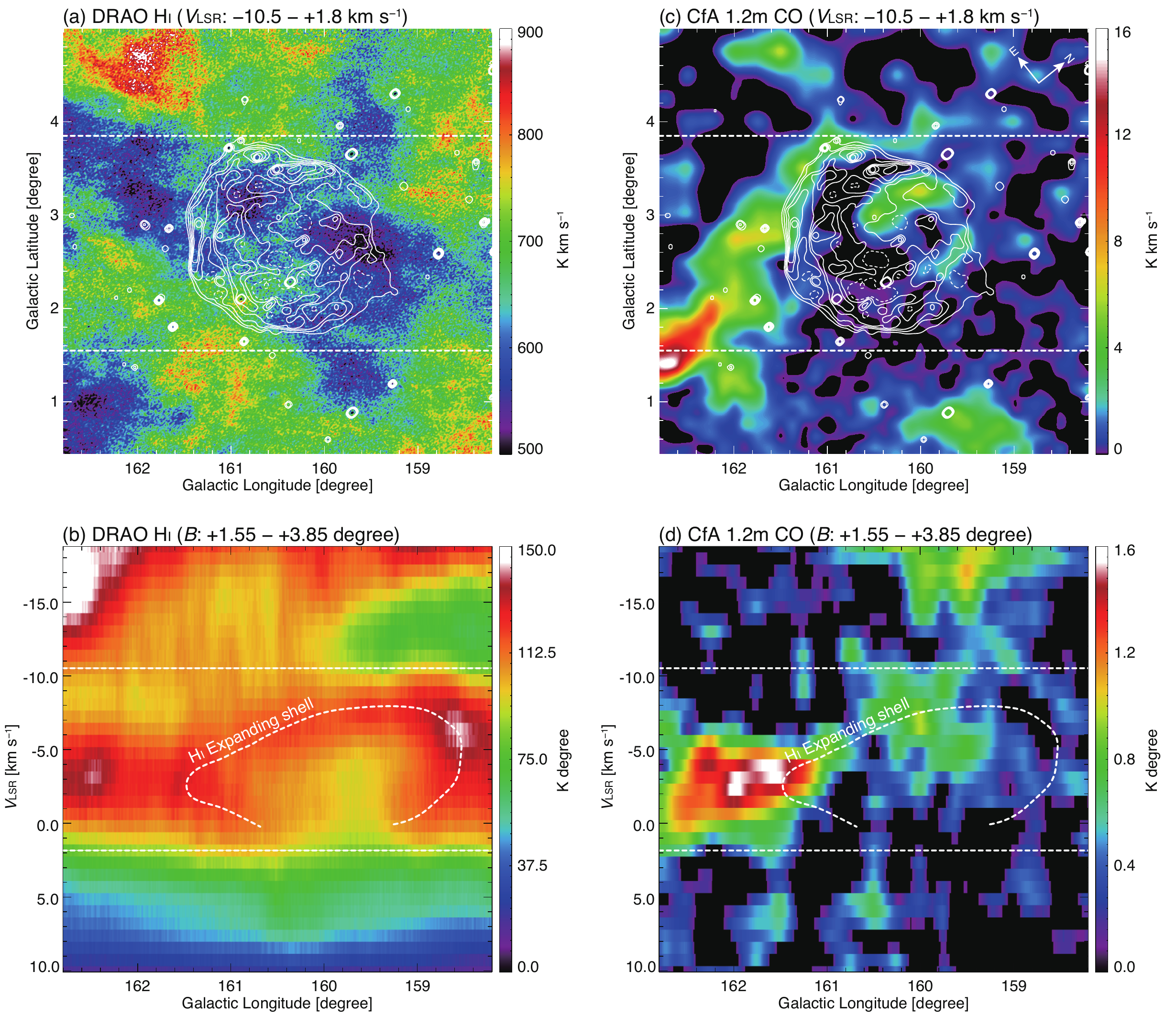}
\caption{(a-b) Intensity distributions of (a) H\,{\sc i} and (b) $^{12}$CO($J$ = 1--0) in the analysis region are shown in Galactic coordinates. The integration velocity of H\,{\sc i} and CO is from $-10.5$ $-$ $+1.8$ km s$^{-1}$. Superposed contours represent the median-filtered radio continuum intensity at 408 MHz. The lowest contour level and contour intervals are 55 Jy beam$^{-1}$ and 5 Jy beam$^{-1}$, respectively. (c-d) Position-velocity diagrams of (c) H\,{\sc i} and (d) CO. The integration range is from $+1.55$ to $+3.85$ degree. Boundaries of the H\,{\sc i} expanding shell are shown by dashed curves.}
\label{Fig4}
\end{figure*}

\begin{figure}
\centering \vspace*{1pt}
\includegraphics[width=0.42\textwidth]{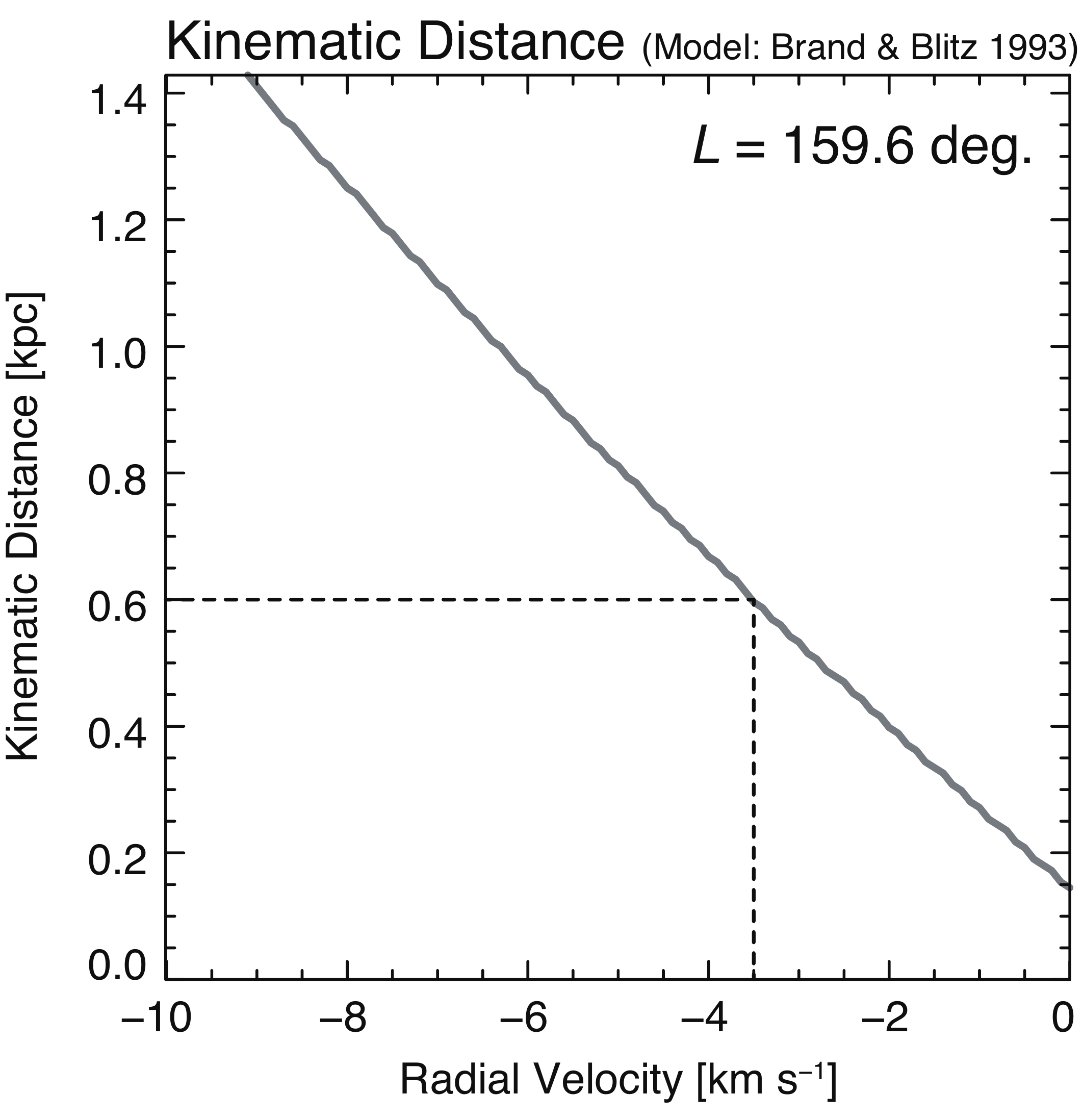}
\caption{Kinematic distance and radial velocity correspondence toward the SNR HB9, assuming the Galactic rotation curve model from \citet{BrBl93}. The dashed lines indicate the central velocity and kinematic distance of the H\,{\sc i} expanding shell.}
\label{Fig5}
\end{figure}

Figs \ref{Fig4}a and \ref{Fig4}b show the distributions of H\,{\sc i} and CO toward the SNR HB9 in the Galactic coordinates. We note that the radio continuum shell at 408 MHz shows a good spatial correspondence with H\,{\sc i} and CO clouds at $V_{\rm LSR}$ = $-10.5$ $-$ $+1.8$ km s$^{-1}$, whose velocity range is roughly consistent with the previous study \citep{LeTi07}. The synchrotron radio shell is nicely along the edge of H\,{\sc i} clouds in north-east and south-west, while the CO clouds trace the eastern shell where the radio continuum is bright. These trends can be understood as the results of magnetic field amplification via the shock-cloud interaction (e.g. \citealp{In09, In12, Sa10, Sa13, Sa17a, Sa17b}), and hence both the H\,{\sc i} and CO clouds likely associated with the SNR. Figs \ref{Fig4}c and 4d show the position-velocity diagrams of H\,{\sc i} and CO. We found a new H\,{\sc i} shell expanding toward the SNR (see the dashed line in Fig. \ref{Fig4}b), whose spatial extent is roughly consistent with that of the SNR. The molecular cloud also shows the expanding motion similar to H\,{\sc i}. This means that the expanding gas motion was possibly created by stellar winds or accretion winds from the progenitor system of the SNR (e.g. \citealt{Sa17b, Sa18, Ku18}), which gives alternative support for the physical association between the SNR shocks and the ISM. The minimum intensity spot corresponding to a geometric center of the expanding shell is $l$ $\sim$ $159\fdg6$ and $V_{\rm LSR}$ $\sim$ $-3.5$ $\pm$ 2.0 km s$^{-1}$, respectively. Based on the Galactic rotation curve model from \citet{BrBl93}, the kinematic distance is calculated as a function of the radial velocity (Fig. \ref{Fig5}). We finally derive the kinematic distance of the H\,{\sc i} expanding shell to be  $\sim$0.6 $\pm$ 0.3 kpc.

\section{Discussion}
\label{discussion}
We have presented the spectral analysis results for HB9 using data from two observation regions of {\it Suzaku}, east and west. We also investigated the morphology and spectral properties of gamma-ray data taken by {\it Fermi}-LAT and analysed archival H\,{\sc i}/CO data of HB9. In the following, we discuss our results.

\subsection{Properties of the X-ray emission from the remnant}
Our spectral analysis shows that the X-ray plasma of both east and west {\it Suzaku} regions is dominated by thermal emission and concentrated below 5 keV. 

We found that the emission of the {\it Suzaku} west region is well reproduced by two thermal plasma, one is CIE plasma and the other is RP. The CIE plasma has solar abundances indicating that the plasma originates from ISM. The RP has slightly enhanced abundance of Si, which suggest the possible presence of SN ejecta.
 
On the other hand, the X-ray spectra of the {\it Suzaku} east region are well explained by two plasma model; a high-temperature in NEI (with $kT_{\rm e}$ $\sim$ 0.97 keV) and a low-temperature in CIE (with $kT_{\rm e}$ $\sim$ 0.51 keV). The elemental abundance of Si in NEI component is slightly higher than the solar values indicating that the plasma is likely to be of ejecta origin.

We found that the electron temperature, $kT_{\rm e}$, is lower in the {\it Suzaku} east region than in the west region. We also found small variations in the absorbing column density, $N_{\rm H}$, which is higher in the {\it Suzaku} east region than in the west region.  

Assuming a distance of 0.6 kpc to the remnant  and using $n_{\rm e}=1.2n_{\rm H}$, we estimated the density of the X-ray emitting gas from the normalization. We found the X-ray emitting gas densities to be $\sim$1.4$f^{-1/2}d_{0.6}^{-1/2}$ ${\rm cm}^{-3}$ and $\sim$0.9$f^{-1/2}d_{0.6}^{-1/2}$ ${\rm cm}^{-3}$ for the {\it Suzaku} east and the west region, respectively, where $f$ is the filling factor. We denoted by $d_{0.6}$ the distance to the source in units of 0.6 kpc. 

We also calculate the recombining age ($t_{\rm rec}=\tau_{\rm rec}/n_{\rm e}$) to be $\sim31\times10^{3}f^{1/2}d_{0.6}^{1/2}$ yr using the best-fitting recombination time-scale $\tau_{\rm rec}$ and normalization listed in Table 2. 

\subsection{Origin of the Recombining Plasma}
In this subsection, we investigate the origin of the RP in the {\it Suzaku} west region. The rapid electron cooling may occur either by the adiabatic expansion (e.g. \citealt{It89, Sh12}) or the thermal conduction from cold clouds (e.g. \citealt{Co99, Sh99}).

In the thermal conduction scenario, if an SNR shock collides with cold MCs, thermal conduction occurs between the SNR plasma and the cloud, the plasma electrons rapidly cool down and the plasma can recombine. But, there is no clear evidence of the SNR-MC interaction for HB9.

In the rarefaction scenario, the plasma cools rapidly when the  shock breaks out of dense circumstellar matter (CSM) into rarefied ISM. We found that the ambient gas density in the {\it Suzaku} west region, which shows an RP, is lower than that in the east region. This result can simply be explained by the rarefaction scenario.

\begin{figure}
\centering \vspace*{1pt}
\includegraphics[width=0.52\textwidth]{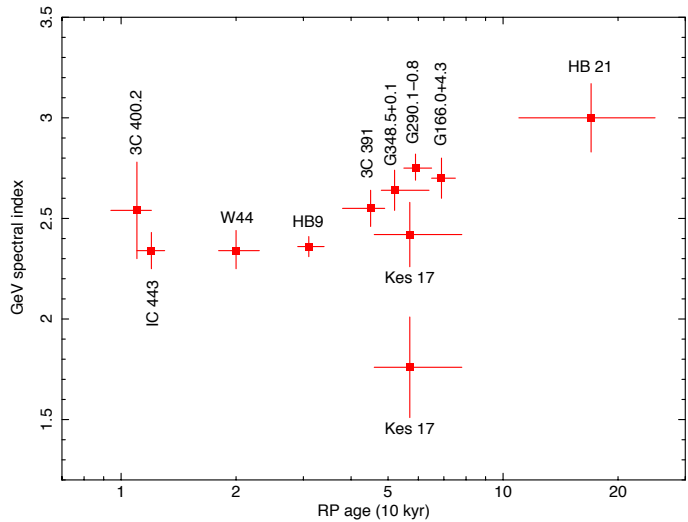}
\caption{RP ages versus GeV spectral indices for recombining GeV SNRs from \citet{Su18} and HB9 from our work.}.
\label{Fig6}
\end{figure}

As already reported by \citet{LeAs95, LeTi07}, {\it ROSAT} and radio observations show that HB9 has a centrally filled X-ray and shell-like radio morphology, similar to that of other MM SNRs (e.g. \citealp{LaSl06, Vi12}). Recently, \citet{Su18} examined the relation between RP ages and GeV spectral indices for nine MM SNRs as shown in their Figure 5. To discuss here HB9 in comparison with MM SNRs showing RPs and gamma-ray emission, we obtain a similar plot including HB9 and illustrate it in Fig. \ref{Fig6}. The GeV index and the RP age of HB9 are following a typical trend-line as seen in other MM-type SNRs (except Kes 17), as given in Fig. \ref{Fig6}. 

\begin{figure*}
\centering \vspace*{1pt}
\includegraphics[width=0.9\textwidth]{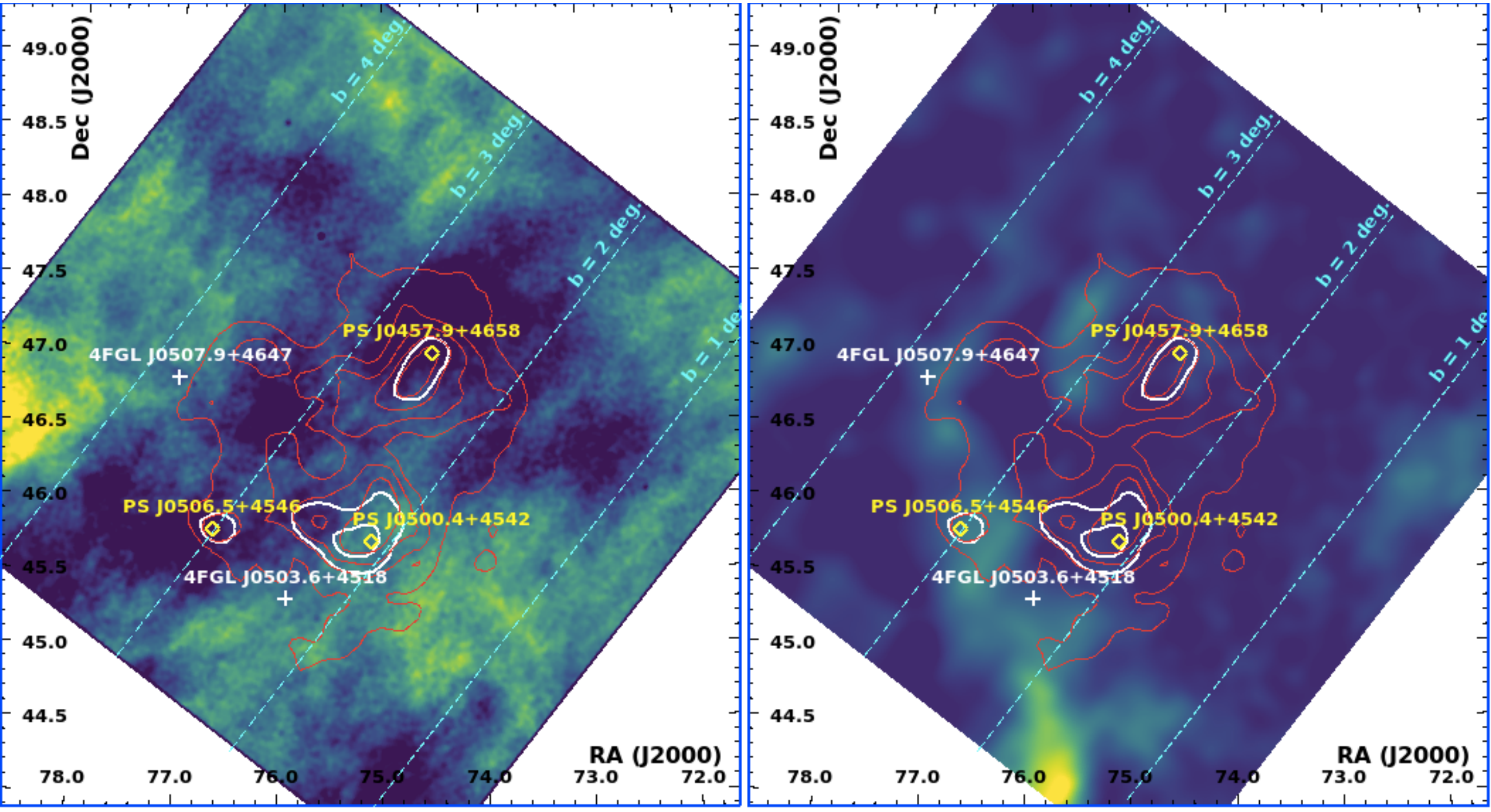}
\caption{ {\bf Left Panel:} The 1.4 GHz H\,{\sc i} line data taken from DRAO that is integrated over the velocity range of [$-10.5$, $+1.8$] km s$^{-1}$ is shown in the range of [500, 900] K km s$^{-1}$. {\bf Right Panel:} The CO intensity from \citet{Da87} integrated over the velocity range of [$-10.5$, $+1.8$] km s$^{-1}$ is shown in the intensity range of 0.0$-$8.0 K km s$^{-1}$. In both panels, the overlaid thin-red gamma-ray contours are produced in the energy range of 0.2$-$300 GeV and they correspond to TS = 25, 49, 64, 81, 100. The bold-white gamma-ray contours are produced in the energy range of 1$-$300 GeV and they correspond to TS = 16,25. Also in both panels, the locations of 4FGL catalog sources are shown with white '+' markers and the three gamma-ray point-source candidates are shown with yellow diamond markers.} 
\label{Fig7}
\end{figure*}

\subsection{Implications for progenitor of SNR HB9}
We examine here the progenitor of HB9 based on the abundance pattern and environment.  

(i) The abundance pattern provides clear signatures of the type of SN explosion: Type Ia SNe producing more iron-group elements, while CC SNe produce a large amount of O, Ne and Mg (e.g. \citealt{Hu95, Vi12}). The {\it Suzaku} X-ray spectra of HB9 reveal line emission from O, Ne, Mg and Si at the soft energies below $\sim$2 keV and a notable absence of iron-group elements in the spectra (see Section 3.1). 

(ii) It is usually assumed that SNRs evolve in denser environments are CC SNRs. As mentioned in Section 3.3, the SNR shell might be close to / expanding into dense material in the north-eastern/eastern side and on the southern side of HB9.

Considering the above properties, we favor a CC origin for HB9 and note that there is no clear evidence of an associated compact object or pulsar wind nebulae detected inside HB9. 

\subsection{Gamma rays \& the atomic and molecular environment}
An extended source analysis of SNR HB9 testing three different extended emission models (Radial Disk, Radial Gaussian and Radio Template models), resulted in The Radio Template model being the best-fit model describing the extension of HB9.

Considering the possibility that the gamma-ray morphology of HB9 could be a result of multiple point-like sources instead of a single extended source, we searched the gamma-ray data for point-sources by excluding the extended emission model of HB9 from the gamma-ray background in the energy range of 1-300 GeV. According to the Akaike information criterion, the best fitting morphological model is where the radio template of HB9 was fit together with the point-source (PS J0506.5+4546). PS J0506.5+4546 is found to be located outside the SNR's shell and its position is not coincident with the pulsar's (PSR B0458+46/PSR J0502+4654) position. 

The spectral parameter $\alpha$ of HB9 measured within the energy range of 200 MeV - 300 GeV was found to be consistent with the spectral results reported by \citet{Ar14} in the same energy range and they agreed with results given in the 4FGL source catalog \citep{Fe19a}, while the $\beta$ parameter was measured slightly lower than the one reported by \citet{Ar14} and \citet{Fe19a}.

Although MM SNRs are usually associated with regions of dense atomic and molecular material, in the case of HB9, no evidence of interaction with MCs has been reported. Nevertheless, we showed on the two bottom panels of Fig. \ref{Fig4} that the shell of the SNR is expanding both into atomic and molecular gas located on the southern, north-eastern and eastern sides of the radio shell of HB9. Therefore, it is possible that gamma rays are produced through the hadronic process through the {\it illuminated-clouds scenario}, where the accelerated high-energy protons escaping from the SNR's shell and penetrating into the densest regions of the ambient matter located on the eastern and southern sides of the remnant result in the production of neutral pions which eventually decay into gamma rays \citep{Ah96,Ya06,Fu09,Oh11}. 

Fig. \ref{Fig7} shows the H\,{\sc i} data obtained by DRAO (left panel) and CO data taken by the CfA 1.2 m telescope \citep{Da87} (right panel) both of which are shown in the velocity range from $-10.5$ km s$^{-1}$ to $+1.8$ km s$^{-1}$. In both of the panels of Fig. \ref{Fig7}, thin-red and bold-white gamma-ray contours indicate the gamma-ray distribution in the energy range of 0.2$-$300 GeV and 1$-$300 GeV, respectively. The bold-white gamma-ray contours overlap spatially with the denser regions of H\,{\sc i}, particularly on the southern part of the SNR's shell, where we see enhancement in the gamma-ray emission (see Fig. \ref{Fig3} bottom-right panel and Fig. \ref{Fig7}). 

Using radio and gamma-ray data, \citet{Ar14} modeled the spectrum to find out the dominating gamma-ray emission scenario. It was concluded that the leptonic scenario, which is the emission of gamma rays from interaction of CMB photons with synchrotron-emitting electrons, is more likely due to the resulting physical parameters being consistent with X-ray measurements. In this scenario, electrons were also assumed to be the source of radio emission seen from the SNR. The bremsstrahlung and hadronic scenarios were found to be less likely due to the fact that they require high density material to interact with. However, in this work we show that the shell of the SNR might be expanding into dense material on the southern and south-eastern/eastern sides of the SNR. This shows the need to repeat the multi-waveband modeling of the spectrum of HB9 using the recent gas and gamma-ray data. This is the topic of the next paper which we are planning.

\section{Conclusions}
\label{conclusions}
In this paper, we have investigated the X-ray and gamma-ray properties of the SNR HB9 based on {\it Suzaku} and {\it Fermi}-LAT observations. We have also presented analysis of both the CO and H\,{\sc i} data to understand the connection between HB9 and its close environment. The main conclusions of our study can be summarized as follows;

\begin{easylist}[itemize]
\vspace{7pt}
& Our X-ray spectral analysis suggests that the plasma of both western and eastern {\it Suzaku} observation regions require two thermal components. We found that the plasma of the western {\it Suzaku} region is in a recombining phase and concluded that the rarefaction scenario is possible explanation for the existence of RP found in this region.

& About 10 years of gamma-ray data were analysed using the most recent (4FGL) Fermi-LAT gamma-ray point source catalog, where HB9 was reported as an extended source. Three extended gamma-ray emission models (Radial Disk, Radial Gaussian and Radio Template) were tested in the 1-300 GeV energy range and the best-fit extended source model was found to be the Radio Template model that uses the 4850 MHz radio continuum map as a spatial model to describe the gamma-ray morphology. 

& We selected the best-fit gamma-ray source model by comparing several different gamma-ray models including the Radio Template model, the 3 Point Sources model, as well as combining the Radio Template model with different number of point-sources. By implementing the Akaike information criterion, it was found out that the Radio Template+1P.S. model best describes the gamma-ray distribution related to HB9. So, in the Radio Template+1P.S. model, the Radio template model represents HB9 as an extended gamma-ray source, while PS J0506.5+4546 is found to be as the point source right outside the extended region of HB9. These two sources are probably physically not related to each other. Moreover, the Radio Template model for HB9 drops the significance of PS J0457.9+4658 and PS J0500.4+4542 below 5$\sigma$ demonstrating that a single extension model is sufficient to characterize the gamma-ray emission morphology of HB9. PS J0506.5+4546 does not spatially correlate with the pulsar PSR B0458+46/PSR J0502+4654.

& From the spectral analysis performed in the energy range of 0.2$-$300 GeV by applying the Radio Template+1P.S. model, HB9 was detected with a significance of $\sim$25$\sigma$ and PS J0506.5+4546 having a power-law type of spectrum was detected with a significance of $\sim$6$\sigma$. The spectral properties of HB9 were found to be in agreement with the results reported by \citet{Ar14} and \citet{Fe19a}. 

& We investigated the archival H\,{\sc i} and CO data, where we detected an expanding shell structure in the velocity range of $-10.5$ and $+1.8$ km s$^{-1}$ that is spatially coinciding with a gamma-ray enhanced region located at the southern rim of HB9. H\,{\sc i} and CO analysis revealed an expanding gas motion, whose spatial extend is roughly correlated with that of the SNR. This motion could be a result of stellar winds or accretion winds of the progenitor system. The distance was found to be $0.6 \pm 0.3$ kpc. The RP emission is detected at a region with relatively lower ambient density.

& We showed in this work that the shell of HB9 might be associated with dense clouds, in particular on the south region of the SNR shell. This association may reveal hadronic gamma rays, although previously it was reported that the leptonic scenario may be dominating over the hadronic gamma-ray emission model \citep{Ar14}. We are going to prepare a second paper including the modeling of the multi-wavelength spectrum in order to understand the contribution of hadronic scenario to the overall gamma-ray emission in HB9. 

\end{easylist}

\section*{Acknowledgments}
We would like to thank all the {\it Suzaku} \& {\it Fermi}-LAT team members. We also would like to thank the anonymous referee for his/her constructive and useful comments. AS is supported by the Scientific and Technological Research Council of Turkey (T\"{U}B\.{I}TAK) through the B\.{I}DEB-2219 fellowship program. This work is supported in part by grant-in-aid from the Ministry of Education, Culture, Sports, Science, and Technology (MEXT) of Japan, No.18H01232(RY), No.15H05694 (HS \& YF), No.16K17664 (HS \& YF) and in part by Aoyama Gakuin University Research Institute. HS is also supported by ``Building of Consortia for the Development of Human Resources in Science and Technology'' of Ministry of Education, Culture, Sports, Science and Technology (MEXT, grant No. 01-M1-0305). The Canadian Galactic Plane Survey (CGPS) is a Canadian project with international partners. The Dominion Radio Astrophysical Observatory is operated as a national facility by the National Research Council of Canada. The CGPS is supported by a grant from the Natural Sciences and Engineering Research Council of Canada.
$~$

{\it {\bf Facility}}: {\it Suzaku}, {\it Fermi}-LAT, {\it ROSAT}, Dominion Radio Astrophysical Observatory, Harvard-Smithsonian Center for Astrophysics 1.2 m MMW-radio Telescope, Green Bank Telescope. 
\vspace{-0.5 cm}

\onecolumn

\twocolumn

\end{document}